# Cell Model of In-cloud Scavenging of Highly Soluble Gases

Alexander Baklanov [a,b], Tov Elperin [c], Andrew Fominykh [c], Boris Krasovitov [c*]

[a] *Sector of Meteorological Model Systems, Research Department, Danish Meteorological Institute Lyngbyvej 100, DK-2100 Copenhagen, Denmark*

[b] *A.M. Obukhov Institute of Atmospheric Physics of the Russian Academy of Sciences, Pyzhevsky 3, 119017, Moscow, Russia*

[c] *Department of Mechanical Engineering, The Pearlstone Center for Aeronautical Engineering Studies, Ben-Gurion University of the Negev, P. O. B. 653, 84105, Israel*

**Abstract**

We investigate mass transfer during absorption of highly soluble gases such as $HNO_3$, $H_2O_2$ by stagnant cloud droplets in the presence of inert admixtures. Thermophysical properties of the gases and liquids are assumed to be constant. Diffusion interactions between droplets, caused by the overlap of depleted of soluble gas regions around the neighboring droplets, are taken into account in the approximation of a cellular model of a gas-droplet suspension whereby a suspension is viewed as a periodic structure consisting of the identical spherical cells with periodic boundary conditions at the cell boundary. Using this model we determined temporal and spatial dependencies of the concentration of the soluble trace gas in a gaseous phase and in a droplet and calculated the dependence of the scavenging coefficient on time. It is shown that scavenging of highly soluble gases by cloud droplets leads to essential decrease of soluble trace gas concentration in the interstitial air. We found that scavenging coefficient for gas absorption by cloud droplets remains constant and sharply decreases only at the final stage of absorption. In the calculations we employed gamma size distribution of cloud droplets. It was shown that despite of the comparable values of Henry's law constants for the hydrogen peroxide ($H_2O_2$) and the nitric acid ($HNO_3$), the nitric acid is scavenged more effectively by cloud than the hydrogen peroxide due to a major affect of the dissociation reaction on $HNO_3$ scavenging.

*Keywords*: In-cloud scavenging, gas absorption, highly soluble gases, scavenging coefficient, atmospheric chemistry

---

[*] Corresponding author Tel.: +972 8 6477067
   Fax: +972 8 6472813
   E-mail: borisk@bgu.ac.il

**Nomenclature**

| | |
|---|---|
| $a$ | radius of a cloud droplet, m |
| $C_{A,G}$ | concentration of a soluble trace gas in a gaseous phase, $\text{mole} \cdot \text{m}^{-3}$ |
| $C_{A,L}$ | concentration of dissolved gas in a droplet, $\text{mole} \cdot \text{m}^{-3}$ |
| $C$ | total concentration of ambient gas, $\text{mole} \cdot \text{m}^{-3}$ |
| $D_{A,G}$ | coefficient of diffusion in a gaseous phase, $\text{m}^2 \cdot \text{s}^{-1}$ |
| $H_A$ | Henry's Law constant, $\text{M} \cdot \text{atm}^{-1}$ |
| $M$ | molar mass, $\text{kg} \cdot \text{mole}^{-1}$ |
| $m = H_A R_g T$ | dimensionless Henry's Law coefficient |
| $r$ | coordinate, m |
| $R$ | radius of a cell, m |
| $R_g$ | universal gas constant, $\text{atm} \cdot \text{M}^{-1} \cdot \text{K}^{-1}$ |
| $q_c$ | flux of a soluble gas, $\text{mole} \cdot \text{m}^{-2} \cdot \text{s}^{-1}$ |
| $t$ | time, s |

*Greek symbols*

| | |
|---|---|
| $\Lambda$ | scavenging coefficient, $\text{s}^{-1}$ |
| $\rho$ | density, $\text{kg} \cdot \text{m}^{-3}$ |
| $\tau = D_{A,L} \cdot t / a^2$ | dimensionless time |
| $\varphi_L$ | volume fraction of droplets in a cloud |

*Subscripts*

| | |
|---|---|
| 0 | initial value |
| A | absorbate (soluble trace gas) |
| G | gaseous phase |
| L | liquid phase |

**1. Introduction**

Wet removal of gaseous pollutants by cloud droplets is involved in various atmospheric processes such as scavenging of gaseous pollutants, cloud processing, acid deposition etc. Heat and mass transfer during gas absorption by liquid droplets is important in various fields of environmental engineering and atmospheric science. Clouds represent an important element in self-cleansing process of the atmosphere (Flossmann, 1998). The consequence for the aerosol climate forcing is that the cooling can be enhanced with increasing atmospheric amount of water-soluble trace gases such as $HNO_3$, counteracting the warming effect of the greenhouse gases (Krämer et al., 2000). New generation of fully online integrated atmospheric chemistry and meteorology models and Earth system models requires more sophisticated and accurate description of interactions of clouds with atmospheric gases and aerosols. Scavenging of the atmospheric gaseous pollutants by cloud droplets is the result of gas absorption mechanism (Pruppacher and Klett, 1997; Flossmann, 1998). Transport of soluble gases in clouds is an integral part of the atmospheric transport of gases and is important for understanding the global distribution pattern of soluble trace gases. Gas scavenging of highly soluble gases by atmospheric water droplets includes absorption of $HNO_3, H_2O_2, H_2SO_4, HCl, NH_3$ and some other gases. The sources of these gases in the atmosphere are briefly reviewed by Seinfeld and Pandis (2006) and Hayden et al. (2008). Soluble gas absorption by droplets without internal circulation was investigated experimentally by Hixson and Scott (1935), Bosworth (1946) and Taniguchi and Asano (1992). Suppression of the internal circulation in falling liquid droplets was achieved by using high-viscosity liquids in the experiments of Hixson and Scott (1935) and Bosworth (1946) and by employing small water droplets in the experiments of Taniguchi and Asano (1992). Hixson and Scott (1935) investigated benzene vapor absorption by liquid droplets of straw oil, and Bosworth (1946) studied $CO_2$ gas absorption by liquid droplets of the concentrated sucrose solution. In these experimental studies droplet diameter was larger than 1mm. Taniguchi and Asano (1992) used water droplets with Sauter mean diameter equal to 0.185 mm, 0.148 mm and 0.137 mm in the experiments with $CO_2$ absorption. Bosworth (1946) compared experimental results with error-function solution of the nonstationary spherically symmetric equation of diffusion for a liquid droplet. Taniguchi and Asano (1992) compared the experimental results with the analytical solution by Newman (see Newman 1931). Dispersed-phase controlled absorption of a pure gas by a stagnant liquid droplet was investigated analytically by Newman (1931), and gas absorption in the presence of inert admixtures when both phases affect mass transfer was analyzed by Clift et al. (1978), pp. 54-55 and by Plocker and Schmidt-Traub (1972). Liquid and gaseous phase controlled mass transfer during soluble gas absorption in the presence of inert admixtures by stagnant liquid droplet was studied also by Chen (2002) by solving the coupled time-dependent diffusion equations for gas and liquid phases. Vesala et al. (2001) solved the general problem of trace gas uptake by droplets under the non-equilibrium conditions numerically and analytically and derived simple formulae for gas uptake coefficient. Chen (2002) analyzed gas absorption by a stagnant liquid aerosol particles using different models, including the perfect absorption model (PAM), rapid diffusion model (RDM), and fully numerical method (FNM). Chen (2002) showed that $CO_2$ and $SO_2$ absorption by aerosol can be approximated well by the RDM while for the analysis of $NH_3$ absorption by aerosol the PAM can be used. Scavenging of soluble gases by single evaporating droplets was analyzed by Elperin et al. (2007, 2008).

In-cloud scavenging of highly-soluble gases was investigated by Levine and Schwartz (1982), Wurzler et al. (1995), Wurzler (1998), Chaumerliac et al. (2000). Levine and Schwartz (1982) assumed that soluble gas scavenging by cloud droplets is governed by physical absorption. By assuming that droplet size distribution in a cloud is determined by empirical distribution Levine and Schwartz (1982) calculated the values of the scavenging coefficient, $\Lambda = 0.2\ s^{-1}$, and the characteristic time for dissolution of gaseous $HNO_3$, $\tau = 5\ s$, in a cloud. The influence of the fine microphysical features such as droplet size distribution and liquid water content on in-cloud scavenging was analyzed also by Chaumerliac et al. (2000). Impact of chemical reactions in a droplet and deviations from the Henry's law on the rate of gas scavenging was studied by Wurzler et al. (1995) and Chaumerliac et al. (2000). Different aspects of scavenging of soluble gases by cloud droplets were discussed also by Mari et al. (2000), Garrett et al. (2006), Djikaev and Tabazadeh (2003), Zhang et al. (2006), Long et al. (2010).

It must be emphasized that in all above mentioned studies mass transfer during gas absorption by cloud droplets was investigated neglecting interaction between the neighboring droplets. This approach allows determining evolution of the concentration of the dissolved gas in a droplet. In the present study we investigate scavenging of highly soluble gas by cloud droplets taking into account diffusion interaction between droplets. Interaction between droplets is accounted for by employing the cell model of the dispersed media (see e.g. Happel, 1958). In this model the individual droplets in the cloud are considered to be located at the center of a unit cell whereas the whole cloud is described by a cellular structure. Such a cell model or parameterizations based on its results can be easily integrated into online coupled meteorology-chemistry or climate-chemistry models, where the cloud processes and chemical transformation of atmospheric pollutants are considered together with two-way interactions. Consequently, in this model it is sufficient to analyze mass transfer in a unit in order to describe mass transfer in a cloud. Cell model approximation allows us to determine the evolution of the concentration of active trace gas in a droplet and in the gaseous phase. In this study we apply cell model of absorption for the analysis of $HNO_3$ and $H_2O_2$ absorption by cloud droplets.

## 2. Mathematical model

*2.1 Governing equations*

The amount of the active gas that can be dissolved in cloud droplets for slightly soluble gases, e.g., $O_2$, $CO_2$ is significantly smaller than the amount of the soluble gas in a cloud. This circumstance allows neglecting interaction between the cloud droplets. For highly soluble gases, e.g., $HNO_3$, $H_2O_2$, $HCl$, the total amount of a soluble gas that can be dissolved in the cloud droplets can be of the same order of magnitude as the total amount of the soluble gas in a cloud. Therefore, gas absorption by cloud droplets results in the overlap of the depleted of soluble gas regions around the droplets, and their diffusion interaction must be taken into account. Interactive absorption by cloud droplets during the uptake of soluble gas can be adequately described in the approximation of a cell model of the dispersed media (see Fig. 1). Assuming that the volumetric liquid water content in the cloud is $\varphi_L$, let us consider a model whereby each droplet is embedded into the gaseous spherical shell with the radius R. This radius is determined from the relation

$$\varphi_L = a^3 / R^3,$$

and a cloud droplet is located at the center of the cell. Assume that the mass fluxes at the cell boundary vanish. The latter condition will apparently be fulfilled at a finite number of the boundary points because in the middle of the straight line connecting the centers of the neighboring cells the mass fluxes to the droplets will be equal in magnitude but opposite in direction (see Elperin and Fominykh 1996, 1999). Consider a spherical droplet with radius "*a*" immersed in a stagnant gaseous mixture. The gaseous mixture contains the inert gas and soluble species that is absorbed into the liquid droplet. In further analysis we assume spherical symmetry. Extending the conditions of vanishing mass flux to the entire cell boundary we arrive at the following set of equations which describe the process.

In the liquid phase, $0 < r < a$:

$$\frac{\partial C_{A,L}}{\partial t} = D_{A,L} \nabla^2 C_{A,L} \qquad (1)$$

In the gaseous phase, $a < r < R$:

$$\frac{\partial C_{A,G}}{\partial t} + v \nabla C_{A,G} = D_{A,G} \nabla^2 C_{A,G}, \qquad (2)$$

where $v$ is gas velocity. It is known that during gas absorption the hydrodynamic flux in the vicinity of gas-liquid interface does not vanish because not all the molecules diffuse into the liquid phase, and certain fraction of them leaves back into the gas (see e.g., Seinfeld 1986, Chapter 6). The origin of this non vanishing hydrodynamic flux is analogous to the origin of the Stefan's flux arising in the vicinity of the evaporating droplets. Let us estimate the magnitude of the gas velocity in the vicinity of gas-liquid interface. Since $\left|\nabla C_{A,G}\right| \sim C_{A,G}/a$ and $\left|\nabla^2 C_{A,G}\right| \sim C_{A,G}/a^2$, the ratio of the convective term to the diffusion term can be estimated as

$$\frac{\left|v \nabla C_{A,G}\right|}{\left|D_{A,G} \nabla^2 C_{A,G}\right|} \sim \frac{v a}{D_{A,G}}.$$

Using the approach suggested by (Elperin et al., 2007) it can be easily showed that molar average velocity at the droplet surface $v_s$ is of the order of

$$v_s \sim \frac{D_{A,G} C_{A,L}}{C^2} \frac{C_{A,G}}{a}.$$

Consequently, the ratio of the convective term to the diffusion term can be estimated as

$$\frac{\left|v \nabla C_{A,G}\right|}{\left|D_{A,G} \nabla^2 C_{A,G}\right|} \sim \frac{\rho_L M_G^2}{M_L \rho_G^2} C_{A,G}.$$

Concentration of $HNO_3$ in a clean troposphere is usually 0.02 – 0.3 ppb and 3 – 50 ppb in a polluted urban air (see e.g., Seinfeld 1986, p. 37). Since for 1 ppb concentration of $HNO_3$ in the atmosphere $C_{A,G}$ is of the order of $10^{-8}$, it can be easily estimated that $\frac{\left|v \nabla C_{A,G}\right|}{\left|D_{A,G} \nabla^2 C_{A,G}\right|} \sim 3 \cdot 10^{-7} \ll 1$.

Consequently, Eq. (2) can be linearized and written as follows:

$$\frac{\partial C_{A,G}}{\partial t} = \frac{D_{A,G}}{r^2} \frac{\partial}{\partial r}\left(r^2 \frac{\partial C_{A,G}}{\partial r}\right), \ a < r < R. \tag{3}$$

The initial conditions for the system of equations (1) and (3) read:

$$t = 0: \ C_{A,L} = 0, \ C_{A,G} = C_0 \tag{4}$$

The conditions of the continuity of mass flux at the gas-liquid interface yields:

$$D_{A,L} \left.\frac{\partial C_{A,L}}{\partial r}\right|_{r=a} = D_{A,G} \left.\frac{\partial C_{A,G}}{\partial r}\right|_{r=a}, \tag{5}$$

It must be noted that the boundary condition (5) is valid in the case when the flux of the molecules which are not absorbed at the droplet surface and leave the surface back into the gas can be neglected.

The boundary condition for the concentration of the absorbate at the surface of the droplet can be obtained from the analysis of the dissociation reactions. The following equilibrium reactions occur when gaseous $HNO_3$ is dissolved in a water droplet (see, e.g., Seinfeld and Pandis 2006):

$$HNO_3\,(g) + H_2O \overset{H_{HNO_3}}{\rightleftharpoons} HNO_3 \cdot H_2O, \qquad H_{HNO_3} = \frac{[HNO_3 \cdot H_2O]}{p_{HNO_3}} \qquad (6)$$

$$HNO_3 \cdot H_2O \overset{K_1}{\rightleftharpoons} H^+ + NO_3^-, \qquad K_1 = \frac{[H^+][NO_3^-]}{[HNO_3 \cdot H_2O]}, \qquad (7)$$

where $HNO_3 \cdot H_2O$ is physically-dissolved $HNO_3$, $NO_3^-$ is the nitrate, $H_{HNO_3}$ is the Henry's constant and $K_1$ is the dissociation constant. For the total concentration of the dissolved nitric acid we obtain the following expression:

$$[HNO_3^T] = [HNO_3 \cdot H_2O] + [NO_3^-]. \qquad (8)$$

Using the Henri law we obtain:

$$[HNO_3 \cdot H_2O] = H_{HNO_3} \cdot p_{HNO_3}. \qquad (9)$$

Equation (9) and dissociation equilibrium equations (7) – (8) yield:

$$[HNO_3^T] = HNO_3^* \cdot p_{HNO_3} = H_{HNO_3} \cdot \left(1 + \frac{K_1}{[H^+]}\right) \cdot p_{HNO_3}, \qquad (10)$$

where $H^*_{HNO_3} = H_{HNO_3} \cdot \left(1 + \frac{K_1}{[H^+]}\right)$ is the effective Henry's law coefficient for the nitric acid. Since $K_1/[H^+] \gg 1$, Eq. (10) implies that

$$[HNO_3^T] = p_{HNO_3} \cdot H_{HNO_3} \cdot K_1/[H^+]. \qquad (11)$$

Taking into account that

$$[NO_3^-] = \frac{K_1 \cdot H_{HNO_3} \cdot p_{HNO_3}}{[H^+]}, \quad [H^+] = \sqrt{K_1 \cdot H_{HNO_3} \cdot p_{HNO_3}} \qquad (12)$$

we find that

$$C_{A,L} = \sqrt{K_{A,0} C_{A,G} \cdot R_g T} \qquad (13)$$

at the surface of a droplet, where $K_{A,0} = H_{HNO_3} \cdot K_1 = 3.3 \cdot 10^6 \exp(-8700(1/298 - 1/T))$ $(M^2 \cdot atm^{-1})$.

The boundary conditions in the center of the droplet and at the cell boundary read:

$$\left. \frac{\partial C_{A,L}}{\partial r} \right|_{r \to 0} = 0, \tag{14}$$

$$\left. \frac{\partial C_{A,G}}{\partial r} \right|_{r=R} = 0 \tag{15}$$

Condition (14) reflects the requirement that the droplet center should be neither a sink nor a source for the absorbate while condition (15) reflects the fact that due to the periodicity condition mass flux at the cell boundary vanishes.

Thus scavenging of soluble trace gas in a cloud is described by the system of equations (1), (3) with the initial and boundary conditions (4), (5), (13) – (15).

*2.2 Method of solution*

The governing equations can be rewritten using the following coordinate transformations:

$$x = 1 - \frac{r}{a}, \tag{16}$$

for the domain $0 \leq r \leq a$, and

$$w = \frac{1}{\sigma}\left(\frac{r}{a} - 1\right) \tag{17}$$

for the domain $a \leq r \leq R$. The parameter $\sigma$ is chosen such that the coordinate w equals 1 at the boundary of the cell. In the transformed computational domains the coordinates, $x$ and $w$, vary in the same range, $x \in [0, 1]$, $w \in [0, 1]$, and can be treated identically in the numerical calculations. The gas-liquid interface is located at $x = w = 0$. In the new coordinates the system of Eqs. (2) - (8) with the initial and boundary conditions reads:

$$\frac{\partial C_{A,L}}{\partial \tau} = \frac{\partial^2 C_{A,L}}{\partial x^2} - \frac{2}{(1-x)} \cdot \frac{\partial C_{A,L}}{\partial x} \tag{18}$$

$$\frac{\partial C_{A,G}}{\partial \tau} = \frac{D_{A,G}}{D_{A,L}} \cdot \frac{1}{\sigma^2}\left[\frac{\partial^2 C_{A,G}}{\partial w^2} + \frac{2\sigma}{(\sigma w + 1)} \cdot \frac{\partial C_{A,G}}{\partial w}\right]. \tag{19}$$

The initial conditions for Eqs. (18) – (19) become

$$\text{At} \quad \tau = 0, \text{ and } \quad 0 < x < 1, \quad C_{A,L} = C_{L,0} \tag{20}$$

$$\text{At} \quad \tau = 0, \text{ and } \quad 0 < w < 1, \quad C_{A,G} = C_{G,0}. \tag{21}$$

The boundary conditions at gas-droplet interface read:

$$-\frac{D_{A,G}}{\sigma} \frac{\partial C_{A,G}}{\partial w}\bigg|_{w=0} = D_{A,L} \cdot \left(1 - \frac{C_{A,G} \cdot R_g T}{p_\infty}\right) \frac{\partial C_{A,L}}{\partial x}\bigg|_{x=0}, \tag{22}$$

where $p_\infty$ is the total pressure in the gaseous phase and

$$C_{A,L} = \sqrt{K_{A,0} C_{A,G} \cdot R_g T}. \tag{23}$$

The boundary conditions at a center of a droplet and at a cell boundary are as follows:

$$\frac{\partial C_{A,L}}{\partial x}\bigg|_{x=1} = 0, \quad \frac{\partial C_{A,G}}{\partial w}\bigg|_{w=1} = 0. \tag{24}$$

The system of parabolic partial differential equations (18)–(19) with the initial conditions (20), (21) and boundary conditions (22) – (24) was solved using the method of lines developed by Sincovec and Madsen (1975). The spatial discretization on a three point stencil was used in order to reduce the system of the time-dependent partial differential equations to the approximating system of coupled ordinary differential equations. In this approach the system of partial parabolic differential equations is approximated by a system of ordinary differential equations in time for the functions $C_L$ and $C_G$ at the mesh points. The mesh points were spaced adaptively using the following formula:

$$x_i = \left(\frac{i-1}{N}\right)^n, \quad i = 1, \ldots, N+1. \tag{25}$$

Equation (25) implies that mesh points cluster near the left boundary where the gradients are steep. In Eq. (25) $N$ is the chosen number of mesh points, $n$ is an integer coefficient (in our calculations $n$ is chosen equal to 3). The resulting system of ordinary differential equations was solved using a backward differentiation method. Generally, in the numerical solution, 151 mesh points and an error tolerance $\sim 10^{-5}$ in time integration were employed. Variable time steps were used to improve the computing accuracy and efficiency.

Taking into account dissociation reactions in a liquid droplet and Eq. (13) we can also obtain the following expressions for the "equilibrium values" of the concentration of the active gas in the gaseous phase and for the total concentration of the nitric acids in the liquid phase:

$$C_{Geq} = \frac{1}{2}\left[\Phi + 2C_{G,0} - \sqrt{\Phi}\sqrt{\Phi + 4C_{G,0}}\right] \qquad (26)$$

$$C_{L,[N(V)]eq} = \frac{1}{\sqrt{2}}\left(K_{A,0}R_g T\left[\Phi + 2C_{G,0} - \sqrt{\Phi}\sqrt{\Phi + 4C_{G,0}}\right]\right)^{1/2} \qquad (27)$$

where $\Phi = \dfrac{K_{A,0}R_g T \varphi_L^2}{1-\varphi_L^2}$

Taking into consideration that $[HNO_3^T] \approx [H^+]$ allows determining pH in a saturated droplet as

$$\mathrm{pH} = -\log\left([H^+] + C_{L,[N(V)]eq}\right). \qquad (28)$$

The equilibrium reactions occurring when the hydrogen peroxide ($H_2O_2$) is dissolved in water are similar to the reactions of dissolution of the nitric acid ($HNO_3$). Hydrogen peroxide dissociates to produce ions $HO_2^-$:

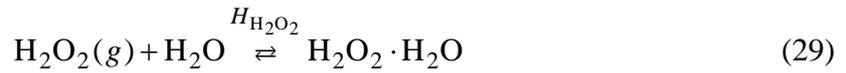

$$H_2O_2(g) + H_2O \underset{}{\overset{H_{H_2O_2}}{\rightleftarrows}} H_2O_2 \cdot H_2O \qquad (29)$$

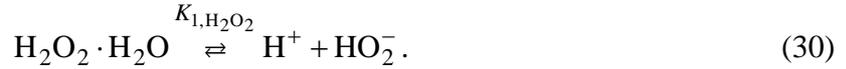

$$H_2O_2 \cdot H_2O \underset{}{\overset{K_{1,H_2O_2}}{\rightleftarrows}} H^+ + HO_2^-. \qquad (30)$$

However in contrast to the nitric acid solution, the hydrogen peroxide solution in water is a weak electrolyte with the dissociation constant $K_{1,H_2O_2} = 2.2 \cdot 10^{-12}$ M at the temperature 298 K. It was shown (see e.g., Seinfeld and Pandis, 2006) that $\dfrac{[HO_2^-]}{[H_2O_2 \cdot H_2O]} = \dfrac{K_{1,H_2O_2}}{[H^+]} < 10^{-4}$ for pH values less than 7.5. Therefore for most atmospheric applications the dissociation of $H_2O_2$ in water can be neglected, and the dissolution of the hydrogen peroxide in water obeys the Henry's law. Consequently, when the hydrogen peroxide absorption by a water droplet is completed, the concentration of the soluble trace gas in the gaseous phase decreases from the initial value $C_{G0}$ down to the "equilibrium value":

$$C_{Geq} = \frac{C_{A,G0}(1-\varphi_L)}{1+\varphi_L(m-1)}. \qquad (31)$$

Accordingly, the concentration of the dissolved gas in the droplet increases from zero up to the "equilibrium value":

$$C_{Leq} = \frac{m \cdot C_{A,G0}(1-\varphi_L)}{1+\varphi_L(m-1)}. \qquad (32)$$

## 3. Results and discussion

The cell model of atmospheric trace gas scavenging by cloud droplets was applied to study the temporal and spatial dependencies of the soluble gas concentration inside the droplets and in a gaseous phase. Recall that in the employed cell model the cloud is viewed as a periodic structure consisting of identical spherical cells with periodic boundary conditions at the cell boundary whereby each cell comprises a water droplet and a spherical gaseous shell. The results of the numerical solution of the system of equations (18) – (19) with the initial and boundary conditions (20) – (24) and for the uniform initial distribution of the soluble trace gas in the gaseous phase are showed in Figs. 2-5. Calculations are performed for absorption of highly soluble gases by water droplets with the radii 10 μm taking into account the dissociation reactions in the liquid phase. The dependencies of the total concentration $[N(V)]$ of the nitric acids and pH in a liquid phase vs. time and radial coordinate are shown in Figs. 2 and 3. Calculations were performed for the ambient temperature 298 K and for the initial concentrations of the nitric acid in a gaseous phase 2 ppb (see Fig. 2) and 50 ppb (see Fig. 3). Inspection of Fig. 2 shows that if the initial concentration of the nitric acid in the gaseous phase is equal to 2 ppb then during gas absorption the acidity of the aqueous solution pH in the droplet varies in the range from 5.5 to 4.1. When the initial concentration of the nitric acid in the gaseous phase is 50 ppb, the acidity of the aqueous solution pH in the droplet varies in the range from 5.5 to 2.7.

The dependencies of the dimensionless concentration of the soluble gas ($HNO_3$) in the gaseous phase vs. time and radial coordinate are shown in Figs. 4 and 5. Calculations were performed for cloud droplets with radii 10 μm immersed into the gaseous phase with the ambient temperature 298 K and the initial concentration of gaseous nitric acid 2 ppb and 50 ppb. Inspection of Figs. 4 and 5 shows that the depleted by soluble gas region around the droplet extends from the surface of the droplet to the boundary of a cell. Figs. 4 and 5 imply that for large time the concentration of $HNO_3$ in the gaseous phase can be determined by Eq. (26).

In Fig. 6 we showed the dependencies of pH in the saturated droplets vs. the initial concentration of $HNO_3$ in the gaseous phase. Calculations were performed using Eqs. (27)–(28) for the ambient temperature 298 K and different values of liquid water content in a cloud. As can be seen from this plot the pH in saturated cloud droplets decreases when the liquid water content decreases.

The dependencies of the average concentration of $HNO_3$ in the gaseous phase, the rate of concentration change $d\bar{c}/dt$ and the scavenging coefficient vs. time are shown in Fig. 7. The curves are plotted for the volume fraction of droplets in a cloud $\varphi_L = 10^{-6}$. This magnitude of the volume fraction of droplets, $\varphi_L = 10^{-6}$, implies that 1 m$^3$ of air contains 1.0 cm$^3$ of water, that is typical for the clouds in the atmosphere (see Seinfeld, 1986, p. 213). The average concentration of HNO$_3$ in the gaseous phase was calculated as follows:

$$\bar{c} = \frac{1}{V_c - V_d} \int C_{A,G}(r) r^2 \sin\vartheta \, dr \, d\vartheta \, d\varphi, \qquad (33)$$

where $V_c$ and $V_d$ are the cell and droplet volumes, correspondingly, $r$, $\vartheta$ and $\varphi$ are the spherical coordinates. The calculations showed in Fig. 7 were performed for cloud droplets with the radii of 5 and 10 μm, correspondingly, and for the fixed volume fraction of droplets $10^{-6}$. We also calculated the scavenging coefficient for the soluble trace gas absorption from the atmosphere:

$$\Lambda = -\frac{1}{\bar{c}} \frac{\partial \bar{c}}{\partial t}. \qquad (34)$$

As can be seen from these plots the scavenging coefficient remains constant, and sharply decreases at the final stage of gas absorption. This assertion implies exponential time decay of the average concentration of the soluble trace gas in the gaseous phase. The latter conclusion can be used for parameterization of gas scavenging by cloud droplets in the atmospheric transport modeling similarly to the aerosol wet deposition (see e.g. Baklanov and Sørensen, 2001). As can be seen from Fig. 7 for the fixed value of volume fraction of droplets equal to $10^{-6}$ the scavenging coefficient in the cloud increases when the droplet radius decreases. This tendency is caused by the increase of the gas-liquid contact surface per unit volume as the droplet radius decreases. We also calculated the rate of concentration change as a function of time (see Fig. 7).

The dependencies of the scavenging coefficient as a function of time for different values of the initial concentration of the soluble trace gas in the gaseous phase are shown in Fig. 8. As can be seen from these plots in the case of the fixed value of the volume fraction of droplets in the cloud and droplet radii, the magnitude of the initial concentration of trace gas in the gaseous phase affects the scavenging coefficient only at the final stage of gas absorption. Notably, when dissociation in water can be neglected (e.g., in the case of H$_2$O$_2$ absorption in a liquid phase) scavenging coefficient is independent of the magnitude of the initial concentration of trace gas in the gaseous phase.

The dependence of the scavenging coefficient for the nitric acid (HNO$_3$) absorbed by cloud droplets vs. time taking into account droplet size distribution in the cloud is shown in Fig.9. In the calculations we assumed the gamma size distribution of cloud droplets with the probability density function:

$$f(a) = \frac{1}{\Gamma(\alpha+1)\beta^{\alpha+1}} a^{\alpha} \exp\left(-\frac{a}{\beta}\right), \tag{35}$$

where $a$ is the radius of the droplet, $\alpha$ is the shape parameter and $\beta$ is the scale parameter. The scale parameter $\beta$ can be determines as follows:

$$\beta = \frac{\bar{a}}{(\alpha+1)} \tag{36}$$

where $\bar{a}$ is the average radius of the droplets and the shape parameter $\alpha = 6$ ( see e.g., Wyser, 1998).

Droplet size distribution was taken into account using the Monte Carlo method whereby the scavenging coefficient was calculated by solving the initial boundary-value problem (1), (3) for a droplet radius that was randomly sampled from the probability density function (35). The scavenging coefficient is determined by averaging the obtained 1000 scavenging coefficients at each time step. Calculations were performed for the average radii 4.7 $\mu$m, that is typical for the Stratocumulus clouds (Sc) and 6 and 7 $\mu$m, which are typical for the Nimbostratus clouds (Ns) (see e.g., Martin et al., 1994). Calculations were performed for different average radii but for the same volume fraction of droplets in a cloud, $\varphi_L = 10^{-6}$. As can be seen from Fig. 9 the scavenging coefficient in the cloud increases when the droplet radius decreases. This tendency is caused by the increase of the gas-liquid contact surface area per unit volume as the droplet radius decreases.

Similar calculations were performed for the hydrogen peroxide $H_2O_2$ scavenged by cloud droplets (see Fig. 10). Calculations were conducted for average radii 4.7 $\mu$m, 6 $\mu$m and 7 $\mu$m and for the constant volume fraction of droplets in a cloud, $\varphi_L = 10^{-6}$.

The "equilibrium fraction" of the total hydrogen peroxide ($H_2O_2$) and nitric acid ($HNO_3$) in the gaseous phase as a function of liquid water content is shown in Fig. 11. These calculations were performed for the ambient temperature 298 K. Here the term "equilibrium fraction" denotes the ratio of the soluble trace gas concentration in the gaseous phase in a state when gas absorption is completed to the initial concentration of the soluble trace gas in the cloud interstitial air. As can be seen from this plot for a cloud liquid water content of 1.0 g/m$^3$ (i.e., $\varphi_L = 10^{-6}$), approximately 40% of the hydrogen peroxide is scavenged by the cloud droplets. Equations (31) – (32) imply that the "equilibrium fraction" of the total $H_2O_2$ does not depend on the initial concentration of $H_2O_2$ in the ambient air. The latter conclusion is the result of the minor role played by dissociation in the process of hydrogen peroxide absorption by cloud droplets.

Despite of the comparable values of the Henry's law constants for the hydrogen peroxide and the nitric acid (see Table 1) the nitric acid is scavenged more effectively by cloud droplets than the hydrogen peroxide (see Fig. 11). The latter assertion results from the major role played by dissociation during absorption of nitric acid by cloud droplets. As can be seen from Fig. 11 the "equilibrium fraction" of the nitric acid depends on the initial concentration of $HNO_3$ in the gaseous phase. However, inspection of this figure shows that for a cloud liquid water content of 1.0 g/m$^3$ only a negligibly small amount of $HNO_3$ remain in the cloud interstitial air even in the case of fairly high atmospheric concentrations of $HNO_3$ in the ambient air e.g., 50 ppb.

**Conclusions**

We accounted for the diffusion interactions between droplets in a cloud caused by the overlap of depleted of soluble gas regions around droplets using the cellular model of a gas-droplet suspension whereby a suspension is viewed as a periodic structure consisting of the identical spherical cells with the periodic boundary conditions at the cell boundary. It is shown that scavenging of highly soluble trace gas in a cloud is described by the system of transient diffusion equations with the corresponding initial and boundary conditions at the droplet center, droplet surface and at the boundary of the cell. The suggested model assumes that gas absorption by cloud droplets is accompanied by the subsequent aqueous-phase equilibrium dissociation reaction. The initial boundary value problem was solved using the method of lines and Monte Carlo simulations. In the calculations we assumed gamma size distribution of cloud droplets. We obtained also the analytical expressions for the "equilibrium values" of concentration of the active gas in a gaseous phase and for the total concentration in the liquid phase for the case of the hydrogen peroxide and nitric acid absorption by cloud droplets. The results obtained in this study allow us to draw the following conclusions:

- Scavenging of highly soluble gases by cloud droplets is described by a system of equations of nonstationary diffusion with the appropriate initial and boundary conditions. Numerical calculations performed for scavenging of the hydrogen peroxide ($H_2O_2$) and nitric acid ($HNO_3$) by cloud droplets allowed us to determine spatial and temporal evolution of the concentration profiles in the droplet and in the interstitial air. The obtained time and space dependent concentration distributions allowed us to calculate the dependence of the scavenging coefficient on time.
- It is shown that scavenging of highly soluble gases by cloud droplets causes a significant decrease of the soluble trace gas concentration in the interstitial air. Calculations conducted for the hydrogen peroxide ($H_2O_2$) and the nitric acid ($HNO_3$) showed that in spite of the comparable values of the Henry's law constants for the hydrogen peroxide and the nitric acid, the nitric acid is scavenged more effectively than the hydrogen peroxide. It is demonstrated that for a cloud with liquid water content of 1.0 g/m$^3$ approximately 40% of $H_2O_2$ is scavenged by cloud droplets while only a negligibly small amount of $HNO_3$ remains in the cloud interstitial air even in the case of fairly high atmospheric concentrations of $HNO_3$ in the ambient air such as 50 ppb. Consequently, the chemical dissociation reaction affects not only the concentration of the dissolved gas in the droplet but also the concentration of the trace soluble gas in the interstitial air.

- Using the suggested cell model we determined the dependencies of the scavenging coefficient as a function of time for different values of the initial concentration of the nitric acid in the gaseous phase. It was found that scavenging coefficient remains constant and sharply decreases only at the final stage of gas absorption. This assertion implies the exponential time decay of the average concentration of the soluble trace gas in the gaseous phase and can be used for the parameterization of gas scavenging by cloud droplets in the atmospheric transport modeling.
- Using the suggested model we calculated temporal evolution of pH in cloud droplets. It was shown that pH strongly depends on the liquid content in the cloud and on the initial concentration of the soluble trace gas in the gaseous phase.

The results of the present study can be useful in an analysis of different meteorology-chemistry models and in particular in various parameterizations of the in-cloud scavenging of the atmospheric soluble gases.

**Acknowledgements**

Table 1 Thermophysical properties of $HNO_3$-water and -$H_2O_2$-water systems

| Gas | $H_A$ [M·atm$^{-1}$] | m=$H_A$·RT For T = 298 K [–] | $D_L$ [m$^2$·s$^{-1}$] | $D_G$ [m$^2$·s$^{-1}$] | $K_1$ [M] |
|---|---|---|---|---|---|
| $HNO_3$ | $2.1·10^5$ | $5.13·10^6$ | $2.6·10^{-9}$ | $1.5·10^{-5}$ | 15.4 |
| $H_2O_2$ | $1.0·10^5$ | $2.44·10^6$ | $2.1·10^{-9}$ | $2.2·10^{-5}$ | $2.2·10^{-12}$ |

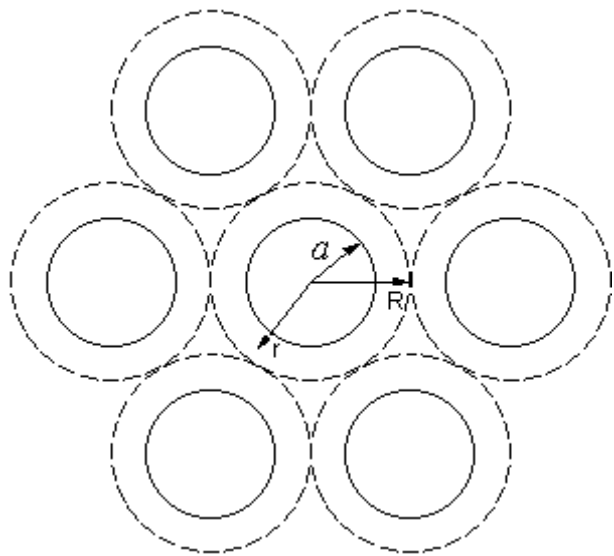

**Fig. 1**. Cell model of gas-droplet media

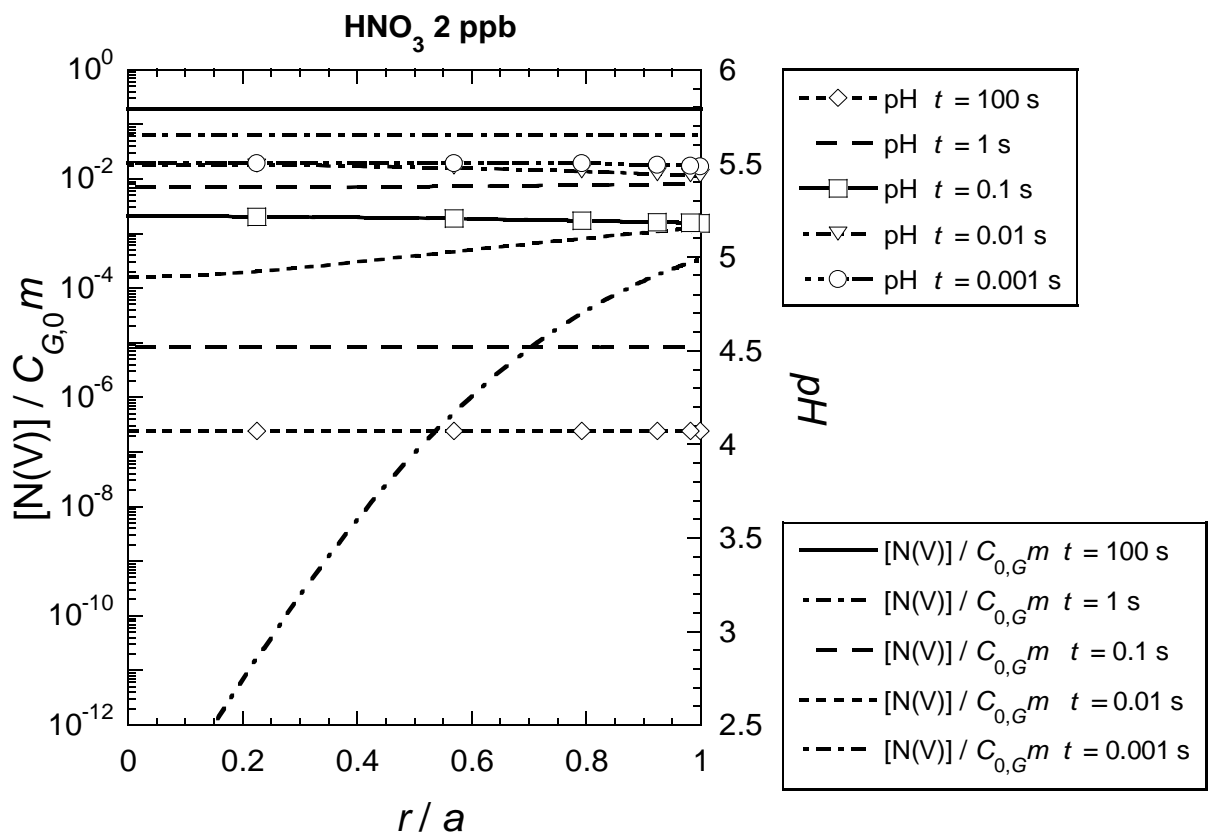

**Fig. 2.** Dependence of the total concentration, $[N(V)]$, of the nitric acids and pH in a liquid phase vs. time and radial coordinate ($C_{G,0} = 2\text{ppb}$).

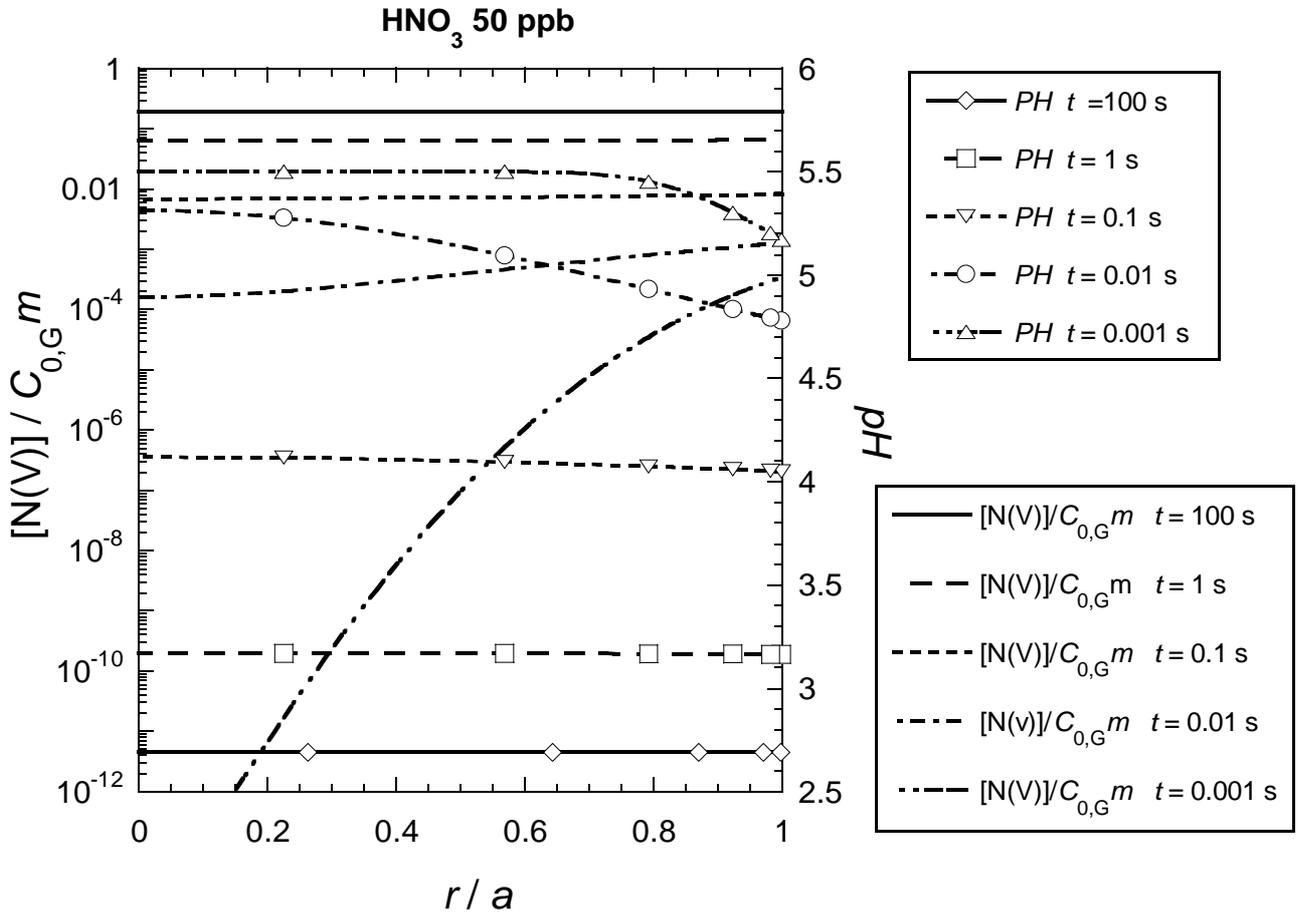

**Fig. 3.** Dependence of the total concentration, $[N(V)]$, of the nitric acids and pH in a liquid phase vs. time and radial coordinate ($C_{G,0} = 50$ ppb).

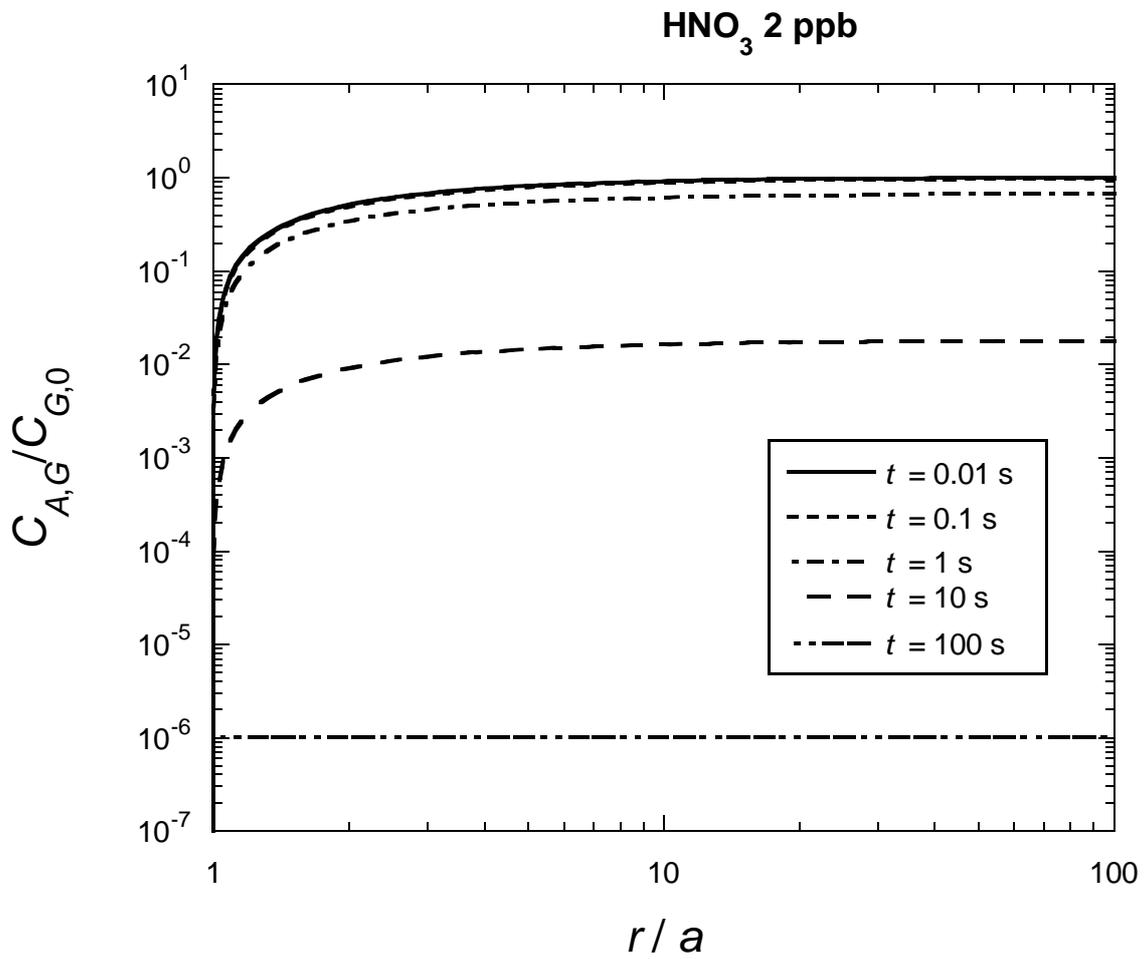

**Fig. 4.** Dependence of the concentration of the soluble gas (HNO$_3$) in the gaseous phase vs. time and radial coordinate ($C_{G,0} = 2$ ppb).

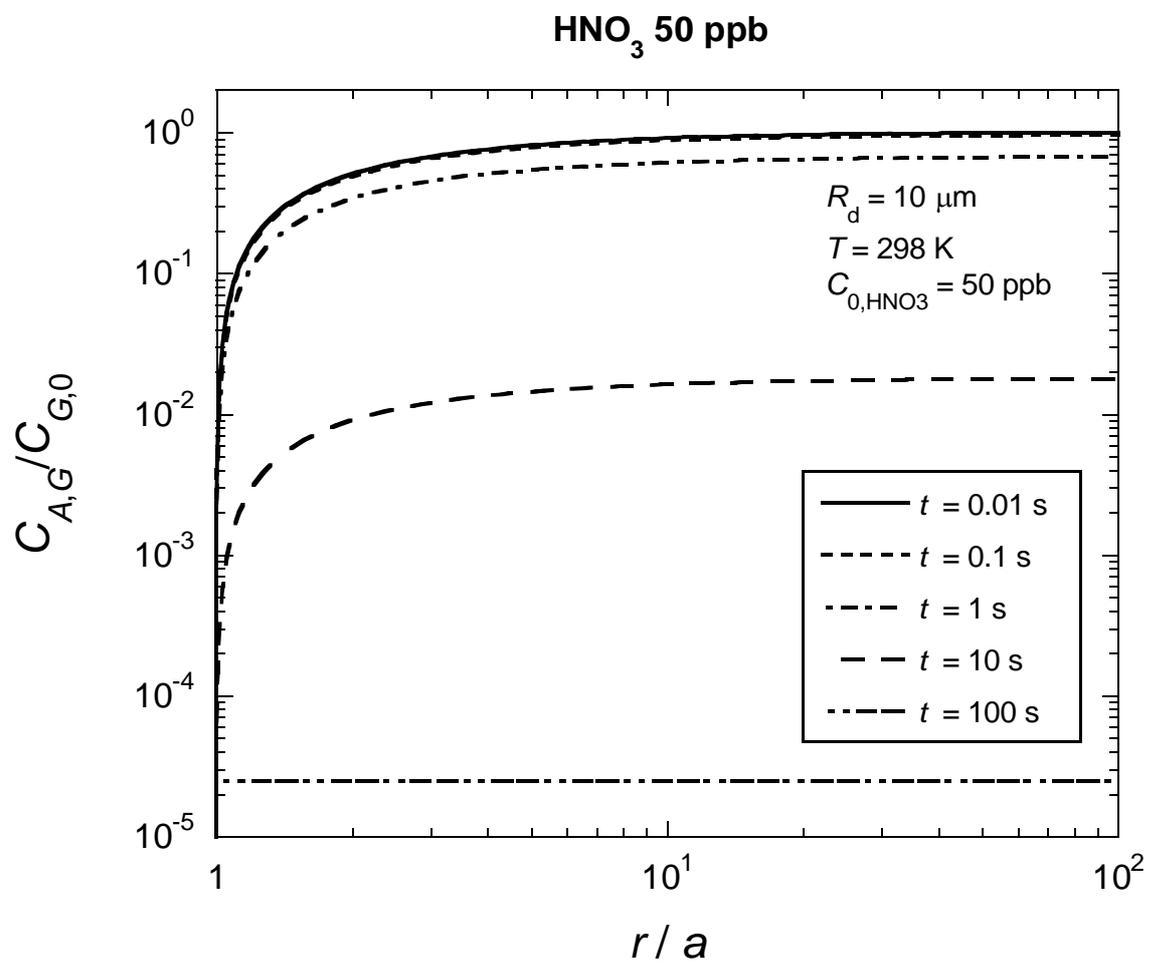

**Fig. 5.** Dependence of the concentration of the soluble gas (HNO$_3$) in the gaseous phase vs. time and radial coordinate ($C_{G,0} = 50$ ppb).

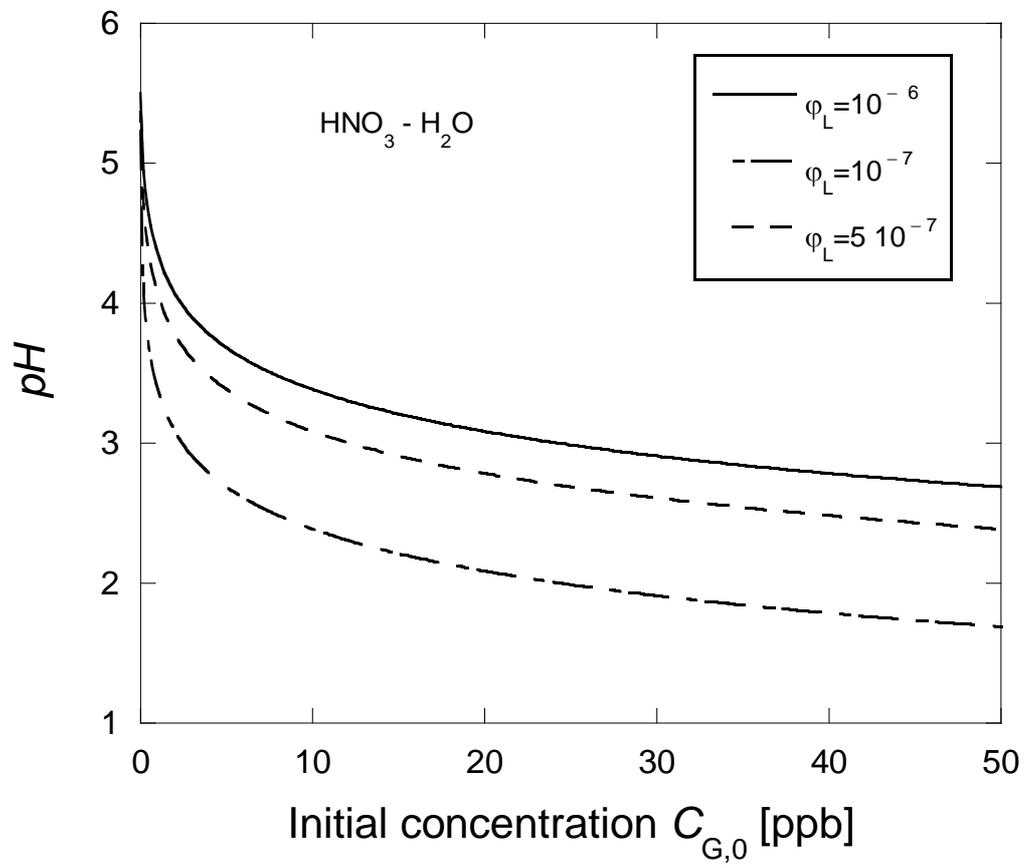

**Fig. 6.** Dependence of pH in the saturated droplet vs. initial concentration of $HNO_3$ in the gaseous phase for different values of liquid water content in a cloud.

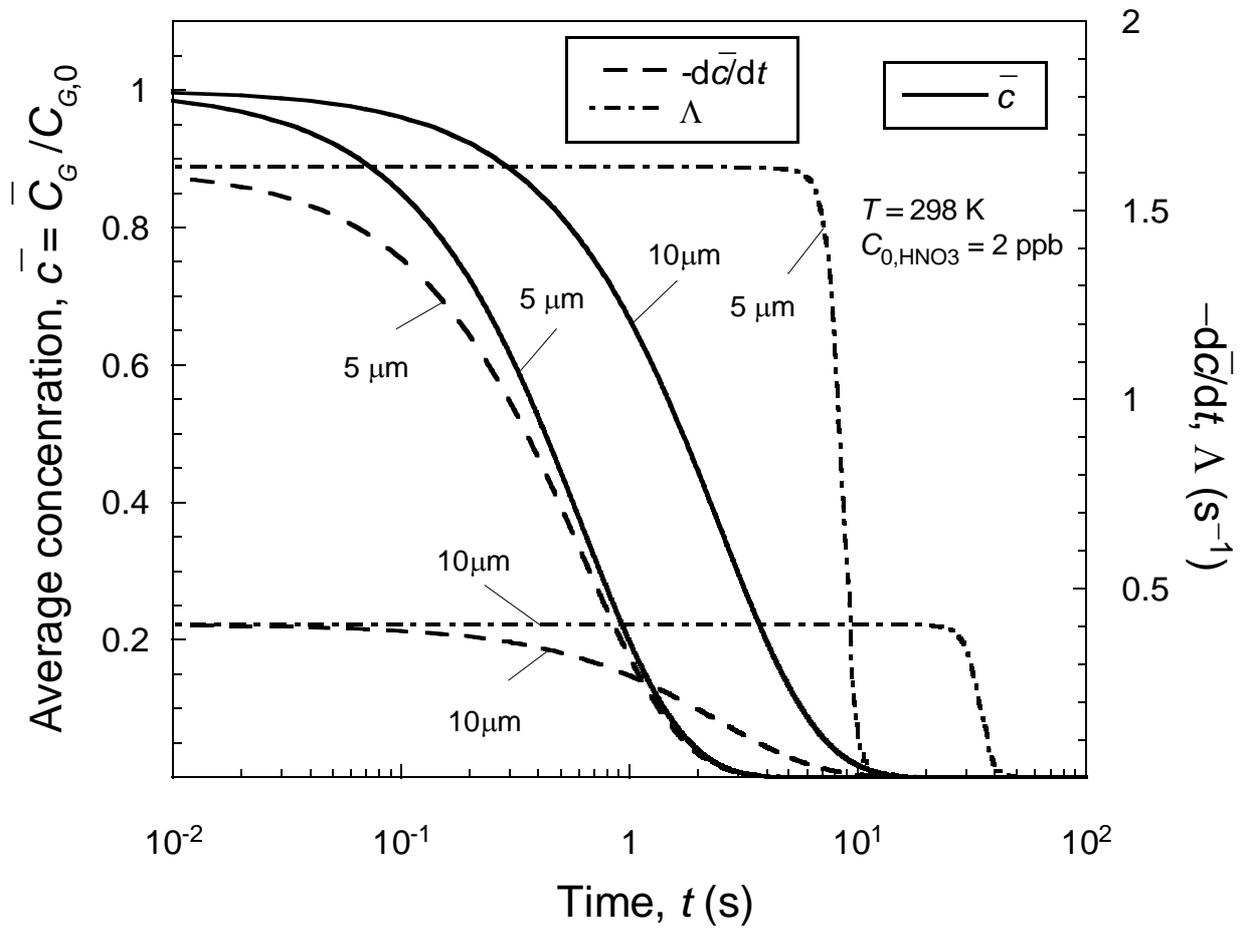

**Fig. 7.** Dependence of the average concentration of $HNO_3$ in the gaseous phase, the rate of concentration change $-d\bar{c}/dt$ and scavenging coefficient vs. time ($\varphi_L = 10^{-6}$).

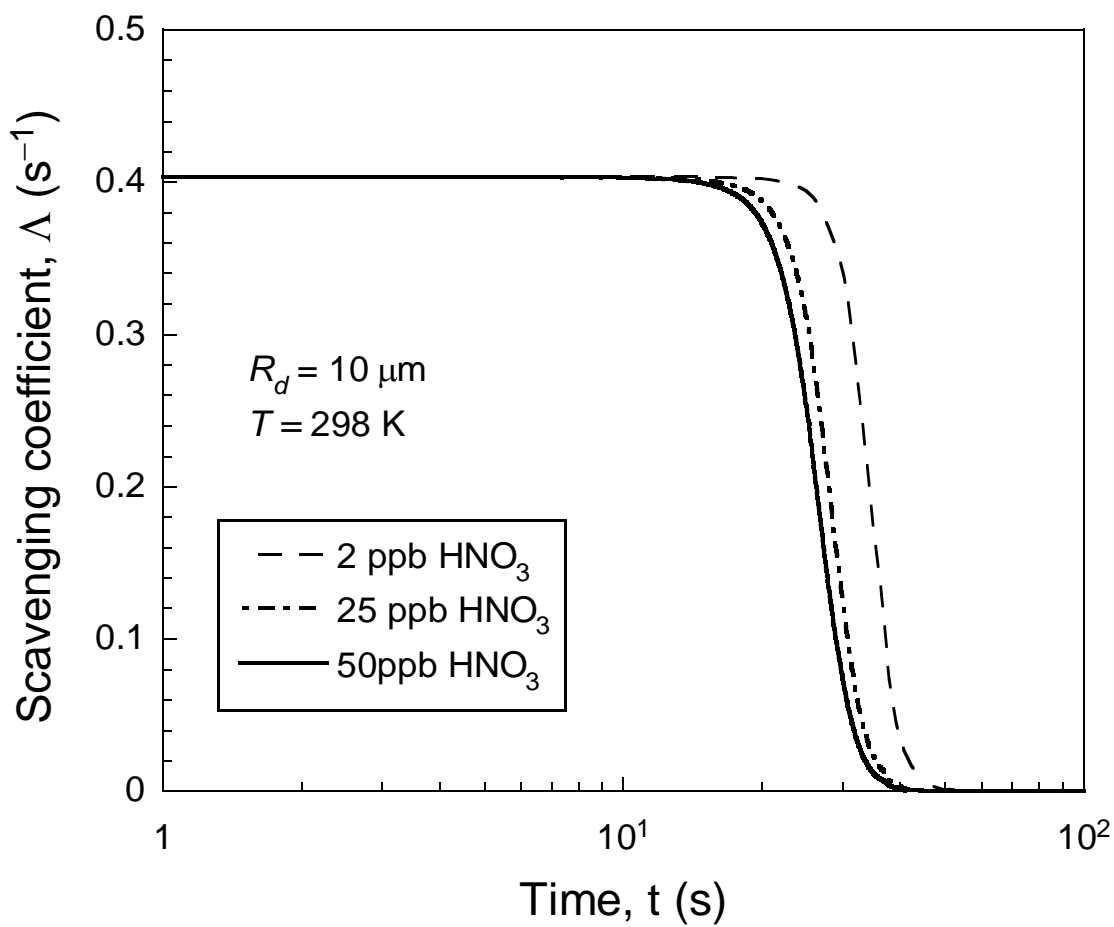

**Fig. 8.** Dependence of the scavenging coefficient of $HNO_3$ vs. time for different values of the initial concentration of soluble trace gas in the gaseous phase.

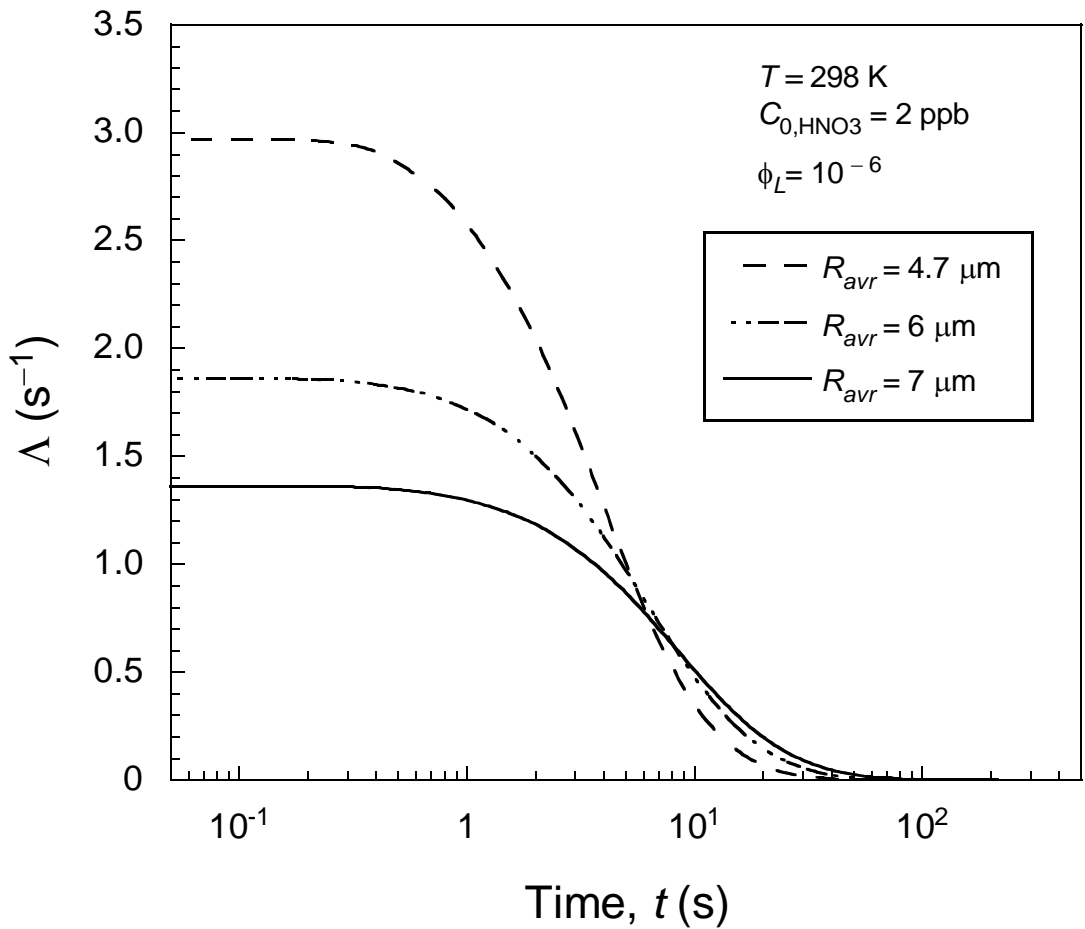

**Fig. 9.** Dependence of the scavenging coefficient for the nitric acid $HNO_3$ on time taking into account gamma droplet size distribution in a cloud.

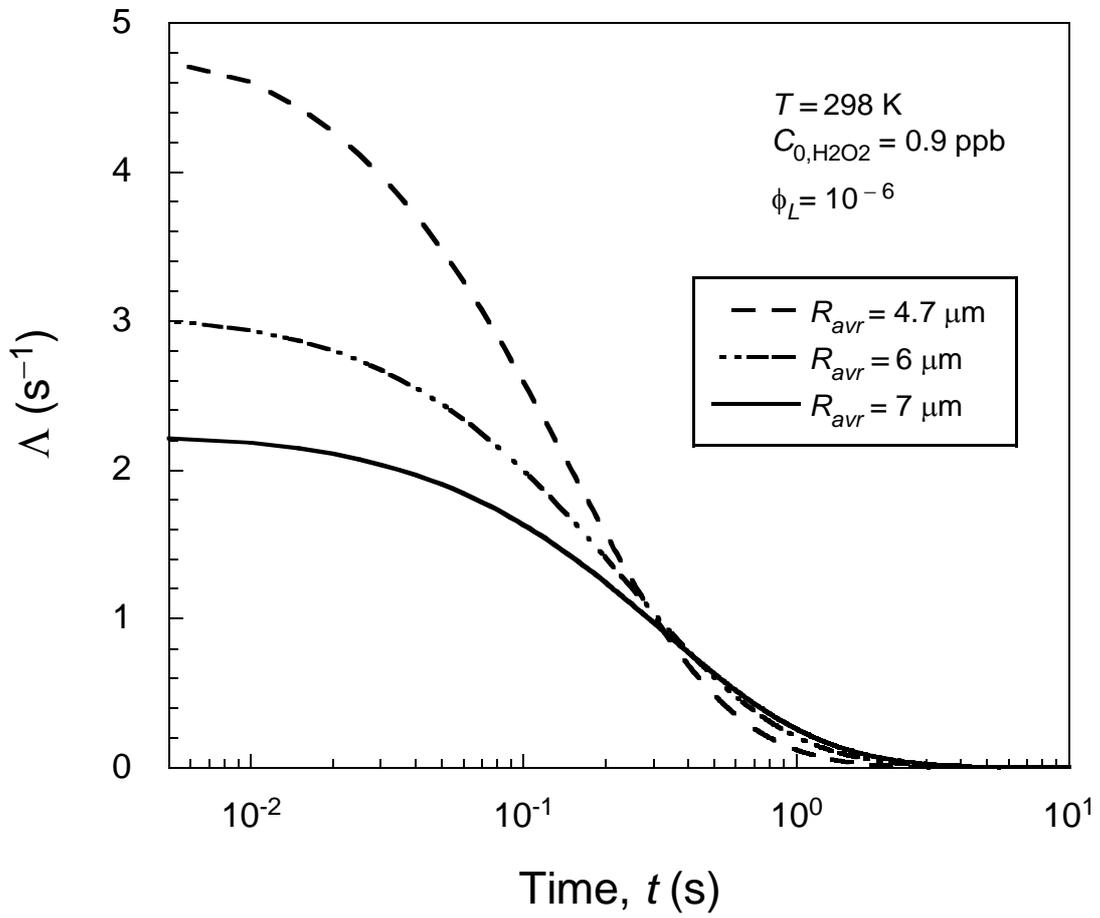

**Fig. 10.** Dependence of scavenging coefficient for the hydrogen peroxide $H_2O_2$ on time taking into account gamma droplet size distribution in a cloud.

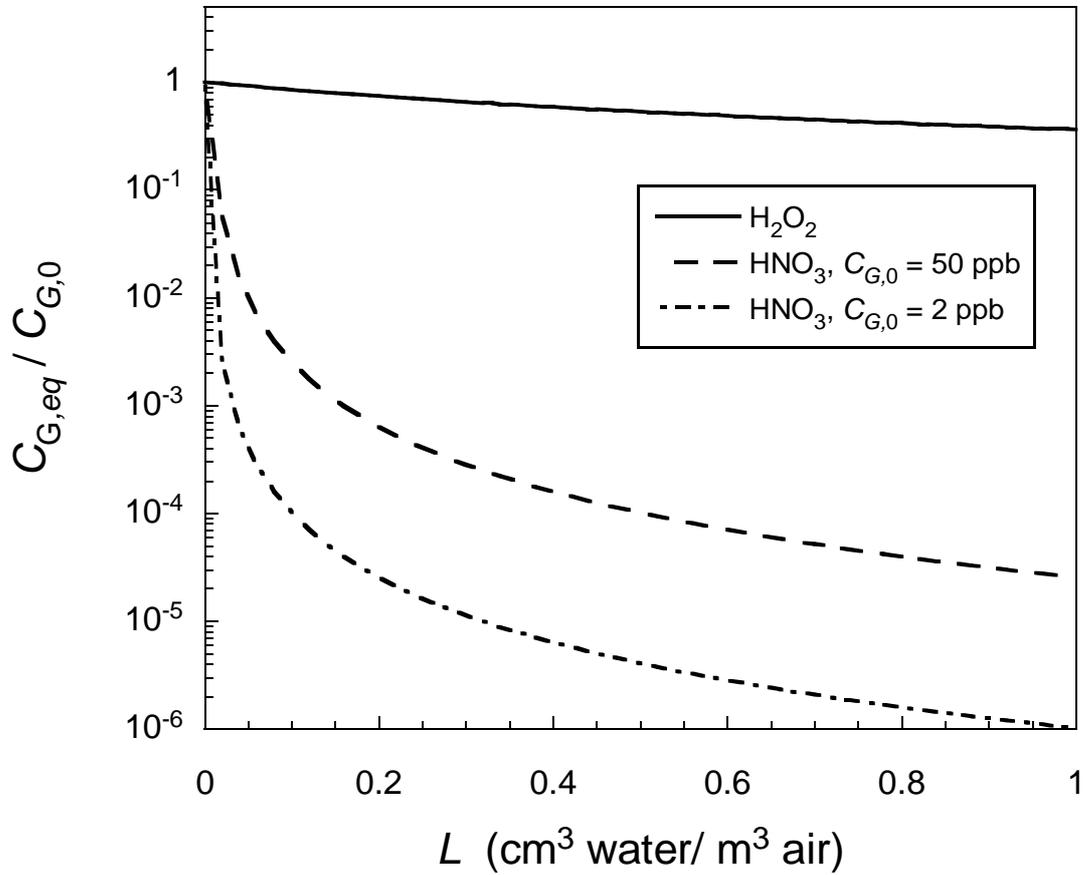

**Fig. 11.** Equilibrium fraction of the total hydrogen peroxide ($H_2O_2$) and nitric acid ($HNO_3$) in the gaseous phase as a function of liquid water content at 298 K.